\documentclass[a4paper]{article}

\usepackage[latin1]{inputenc}
\usepackage[english]{babel}
\usepackage[T1]{fontenc}
\usepackage{graphicx}
\usepackage{bm}
\usepackage{setspace}

\usepackage{amsmath,amsthm,amssymb}

\usepackage[sc]{mathpazo} 
\usepackage[T1]{fontenc} 
\usepackage{microtype} 

\usepackage[hmarginratio=1:1,top=32mm,columnsep=20pt]{geometry} 

\usepackage[hang, small,labelfont=bf,up,textfont=it,up]{caption} 
\usepackage{booktabs} 
\usepackage{float} 
\usepackage{hyperref} 

\usepackage{lettrine} 
\usepackage{paralist} 

\usepackage{abstract} 

\usepackage{titlesec} 

\titleformat{\section}[block]{\large\scshape\centering}{\thesection.}{1em}{} 
\titleformat{\subsection}[block]{\large}{\thesubsection.}{1em}{} 

\usepackage{fancyhdr} 
\numberwithin{equation}{section}
\newcommand{\zeroped}[1]{{\scriptscriptstyle #1}}

 \date{}

\title{\vspace{-20mm}\fontsize{14pt}{15pt}\selectfont\textbf{ Time dependent friction in a free gas }} 

\author{
 \large  
\, Cristiano Fanelli\footnote{Massachusetts Institute of Technology, Laboratory for Nuclear Science, Cambridge, MA 02139, USA. Email: cfanelli@mit.edu},   \, 
Francesco Sisti\footnote{ Dip. di Scienze di Base e Applicate per l'Ingegneria, ''Sapienza'' Universit\'a di Roma , Via A. Scarpa 16, 00161 Roma, Italy. Email: francesco.sisti@sbai.uniroma1.it}, 
 \, Gabriele V. Stagno \footnote{Dip. di Fisica, ''Sapienza'' Universit\'a di Roma , Piazzale Aldo Moro 2, 00185 Roma, Italy. Email: gabriele.stagno@uniroma1.it}     \vspace{-5mm}
 }

\begin{document}

\maketitle

\begin{abstract}

We consider a body immersed in a perfect gas, moving under the action of a constant force $E$ along the 
$ x $ axis . 
We assume the gas to be described by the mean-field approximation and interacting elastically with the body, we study the friction exerted by the gas on the body fixed at constant velocities.
The dynamic in this setting was studied in \cite{nostro}, \cite{E=0} and \cite{CONVEX}
for object with simple shape, the first study where a simple kind of concavity was considered was in \cite{CILINDRO}, showing new features in the dynamic but not in the friction term.
The case of more general shape of the body was left out for further difficulties, we believe indeed that there are actually non trivial issues to be faced for these more general cases.

To show this and in the spirit of  getting a more realistic perspective in the study of friction problems, in this paper we focused our attention on the friction term itself, studying its behavior on a body  with a more general kind of concavity and fixed at constant velocities. 
We derive the expression of the friction term for constant velocities, we show how it is time dependent and we give its exact estimate in time. Finally we use this result to show  the absence of a stationary velocity in the actual dynamic of such a body. \\
\\
\emph{Keywords : Viscous friction; microscopic dynamics. }
\\
\\
  \emph{AMS Subject Classification: 70F40, 70K99, 34C11}
\end{abstract}

\tableofcontents 
\section{Introduction}

We consider a body moving at a constant velocity in in a homogeneous fluid at rest, under the action of a constant force $E$, The gas is initially at thermal equilibrium and it interacts through elastic collisions with the body.
The aim is to show the effect of the body's shape to the friction acting on it, giving an exact estimate of it.

In \cite{nostro} a model of free gas of light particles elastically interacting with a simple shaped body (a stick in two dimensions or a disk in three dimensions) is studied  and it is proven that the system possesses a stationary velocity  
$ V_{\infty} $ which is also the limiting velocity and that the asymptotic time behavior in approaching this $ V_{\infty} $ is power-law.
More precisely, assuming the initial velocity $ V_0 $ such that $ V_{\infty}-V_0 $ is positive and small, it was proven that:
\begin{equation}
\label{intro trend d+2}
| V_{\infty}-V(t)| \approx \frac{C}{t^{d+2}}\, ,
\end{equation}
where $ d=1 $, 2, 3 is the dimension of the physical space and $ C $ is a constant, depending on the medium and on the shape of the obstacle.
This result, surprising for not being exponential, is due to re-collisions that can occur between gas particles and the body while it is accelerating.

Similar model (\cite{diffusivo} )  have been studied where a stochastic kind of interaction between the gas and the body is assumed: when a particle of the medium hits the body it is  absorbed and immediately stochastically re-emitted with a Maxwellian distribution centered  around the body velocity, in this case the behaviour was found to be $ O( \frac{1}{ t^{ d+1} } ) $


In \cite{NUMERICO 1} and \cite{NUMERICO 2} these models have been numerically studied, in particular they computed the dynamic with stochastic interaction carried out in \cite{diffusivo} for a disk subjected to an harmonic force confirming the analytical results (for analytical studies see \cite{E=0} and references quoted therein), here they removed the hypothesis of initial velocity close to $ V_{\infty} $,   showing that this doesn't affect the dynamic.

The question of whether the trend of the solution was connected with the simple shape chosen for the object in the above mentioned articles  was firstly  faced in \cite{CONVEX}, in the domain of elastic collisions, where it was studied the evolution of a general convex body, and  it was confirmed the  same power decay expressed in (\ref{intro trend d+2});  this was mainly due to an important feature that the convex body  shares with the first case of a simple disk, namely in both cases colliding gas particles bounce away from the body.

In \cite{CILINDRO}  the hypothesis of convexity was then removed and it was studied the case of a body with lateral barriers of finite length parallel to the axis of motion (a box-like object in two dimensions), it was shown that this simple kind of concavity significantly altered the asymptotic of the system; it was the first case in which there was a change in the power of decay: $ O(  t^{-3} ) $  and no change from two to three dimensions.
The case of more general shape of the body was left out for further difficulties, we believe indeed that there are actually non trivial issues to be faced for these more general cases.

To show this and in the in the spirit of  getting a more realistic perspective in the study of friction problems, in this paper we studied the friction term itself  acting on a body fixed at constant velocities but with a more general kind of concavity, for example the friction exerted by the air on a vehicle moving at  constant velocity, ( as a practical image one can think to the air on the windshield of a car moving at constant speed)
We show atypical features, namely the presence of a time dependent friction for constant velocities.

In all previous cases in fact  the object (the disk, the convex body or  the box) fixed at any constant velocity experienced a constant friction.
The reason being that a gas particle that hits a disk or a convex body moving at a fixed velocity cannot undergo a second collision with it, thus canceling the effect of recollisions and possible time dependencies in the friction,
as far as the box shaped object in \cite{CILINDRO},  although recolisions can occur even when it is fixed at constant velocity, they will necessarily occur with the lateral barriers of the body which are parallel to the axis of motion, therefore the momentum transferred along the motion direction is zero, thus canceling again recollisions contributions. 

This is not true anymore if we consider the case of the present paper were a slightly more general kind of concavity is considered, namely a body with tilted lateral barriers (see Section \ref{Section our problem} ), these trap particles which, bouncing  inside, correlate through times different areas of the inner side of the body.
The time dependence in the friction term is vanishing and we give an exact estimate of it. \\
We then employ this result to prove  that in the actual dynamic of the body (that is a body which is not fixed in velocity) there can't be a stationary velocity.
The gas is assumed to be made of free particles (see \cite{CERCIGNANI} on Knudsen gas) elastically interacting with the body and it is studied in the mean field approximation.

\section{The model }

\label{Section our problem}

We consider for sake of simplicity the two dimensional case.
In particular we  consider a symmetric angle-shaped body moving towards the $ x $ direction in a Vlasov gas, the body is constrained to stay with its base orthogonal to the x axis, with the center moving along the same axis and with the hollow base facing forwards, the thickness of the body is assumed negligible for sake of simplicity.
The system is immersed in a perfect gas in equilibrium at temperature $T$ and with constant density $ \rho $, assumed in the mean field approximation. (that is the limit in which the mass of the particles goes to zero,while the number of particles per unit volume diverges, so that the mass density stays finite.)
We will study the force exerted by the gas on to the body moving at a fixed velocity.
We refer to our body  as $ C(t) =  C_+(t) \bigcup C_{-}(t) $,
\begin{equation}
\label{c+-}
C_{\pm}(t)= \{ (x, y) \in \mathbb{R}^2 : \, x= X(t) +\eta \cos(\theta) ; \quad y= \pm \eta \sin(\theta) \, , \, \eta \in [0,L]  \}
\end{equation}
where $ \theta \in  [\frac{\pi}{4}, \frac{\pi}{2}) $ represents the angle between the $ x $ axis and the upper side of the body whose vertex is $ \bm{O}(t) =(X(t), 0). $\\
We refer to $ \bm{\hat{N}}(\bm x) $, $ \bm x \in C(t) $ as the right unit normal vector to $C(t)$, namely $ \bm{\hat{N}}(\bm x) \cdot \bm{\hat{x}} > 0$; besides for further convenience we set $ \bm{\hat{n}} = \bm{\hat{N}}(\bm x):\bm x \in C^+ (t)$, and $ \bm{\hat{p}} = \bm{\hat{N}}(\bm x): \bm x \in C^- (t)$, in particular $ \bm{\hat{n} }  =( \sin(\theta), -\cos(\theta ) ) $ and $ \bm{\hat{p} }  =( \sin(\theta), \cos(\theta ) ) $ (see Figure 1).


Let then $ f(\bm{x},\bm{v}, t) , \ (\bm{x},\bm{v}) \in \mathbb{R}^2 \times \mathbb{R}^2 $ 
be the mass density in the phase space of the gas particles.
It evolves according to the free Vlasov equation:
\begin{equation}
\label{vlasov}
(\partial_t +\bm{v} \cdot \nabla_{\bm{x}}) f(\bm{x},\bm{v}, t) =0, \qquad \bm{x} \notin C(t)
\end{equation}

Together with eq. (\ref{vlasov}) we consider the boundary conditions.
They express conservation of density along trajectories with elastic reflection on $ C(t)$. 
Let then $\bm{v}$ be the velocity of a gas particle that hits the body at time $t$ at the collision point $R \in C(t)$ and  $ V(t) = \dot{X}(t)$ the velocity of the body, denoting by $v_N= \bm{v}  \cdot \bm{\hat{N}} $  and by  $V_N = V(t) \bm{\hat{N}}  \cdot \bm{\hat{x}}$, imposing elastic reflection in $R$, we have for the outgoing velocity $\bm{w}$ (for a derivation see  the appendix):
\begin{equation}
w_{N} = 2V_N(t)  - v_N, \quad w_{N_{\perp}} = v_{N_{\perp}} 
\end{equation}
where $ v_{N_{\perp}}=  \bm{v}- v_N \bm{\hat{N}}  $
\\
The boundary condition then reads as follows:
\begin{equation}
\label{boundary conditions}
f_+ (\bm{x},\bm{w}, t) = f_- (\bm{x},\bm{v}, t) \qquad \bm{x} \in C(t) \\
\end{equation} 
where
\begin{equation}
f_{\pm}(\bm{x},\bm{v}, t)= \lim _{\epsilon \to 0^+ } f(\bm{x} \pm \epsilon \bm{v}, \bm{v},t \pm \epsilon).
\end{equation}

\begin{figure}
\center{
\includegraphics[width=9cm]{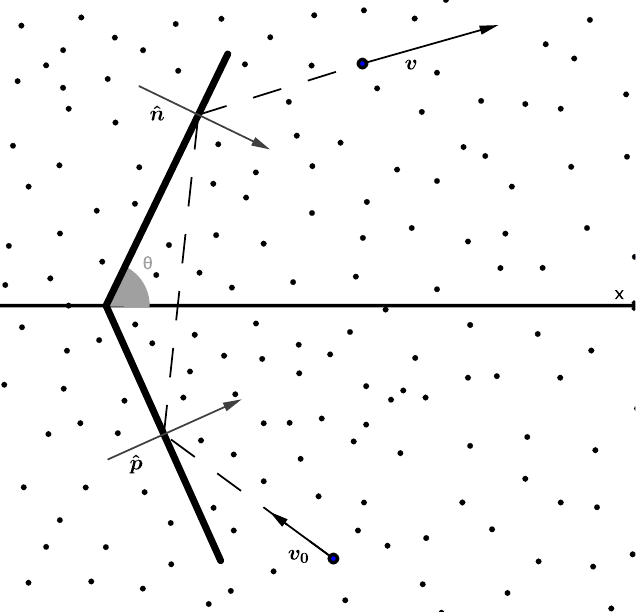}
\caption{ The body  moving in the gas; the dashed line represents the trajectory of one gas particle hitting the body}
}
\end{figure}

Finally we give the initial state of the gas, assumed in thermal equilibrium through the Maxwell- Boltzmann distribution
\begin{equation}
\label{distribuzione di Maxwell} 
f(\bm{x}, \bm{v},0)= f_0(\bm{v}^2)= \rho ( \frac{\beta}{\pi} )  e^{- \beta \bm{v}^2} 
\end{equation}
with $\beta=\frac{1}{k_B T}$.

The above equation for the gas are clearly coupled with those of the body immersed in it, which are:
\begin{subequations}
\label{d/dt V = E- F(t)}
\begin{gather}
\frac{d}{dt} X(t) = V(t), \\
\frac{d}{dt} V(t) = E - F(t), \\
X(0) = 0, \qquad V(0) = V_0, 
\end{gather} 
\end{subequations}
where $ E $ is the constant external force acting along the x-axis and
\begin{equation}
\label{F}
\begin{split}
F(t)= 4 \sin(\theta)  \int \limits _ { 0 }^L  d \eta \int \limits _{ \bm{v}' \cdot \bm{\hat{n}}< 0 } d \bm{v}
( \bm{v}' \cdot \bm{n}  )^2  f_-(X(t) +\eta \cos(\theta),\eta \sin(\theta) ,\bm{v},t)  +\\
- 4 \sin(\theta)  \int \limits _ { 0 }^L d \eta \int \limits _{ \bm{v}' \cdot \bm{\hat{n}} > 0 } d \bm{v}
( \bm{v}' \cdot \bm{\hat{n}}  )^2 f_-(X(t) +\eta \cos(\theta),\eta \sin(\theta) ,\bm{v},t)
\end{split}
\end{equation}
is the action of the gas on the disk,  
being $ \bm{v}'  = \bm{v} -V(t) \bm{\hat{x}} $ the velocity in the reference system of the body,
and where $  \bm{\hat{n} }  = \bm{\hat{N}_R}  $ for $ \bm{x} \in C_+(t) $, in particular 
$ \bm{\hat{n} }  =( \sin(\theta), -\cos(\theta ) ) $.

We give here a derivation of eq.(\ref{d/dt V = E- F(t)}) and (\ref{F}). \\
The contributions to $ F(t) $ coming from $  C_+(t)  $ and $  C_{-}(t) $ are the same for the symmetry on the $ y $ axis of the gas distribution and the shape of the body, we thus  consider the action of the gas on $  C_+(t)  $.
Our body, while moving, is subjected to multiples collision with gas particles, if we write its total variation of momentum in an interval $(t,t + \Delta t )$ along the $x$ axis as $ \Delta V(t)  $ and the variation of momentum due to collisions with gas particles as $ \Delta V_{coll}(t)  $, we have that ($M$ is the mass of the body) 
\begin{equation}
M \Delta V(t) = E \Delta t +  M \Delta V_{coll}(t).
\end{equation}

In the following for any vector $ \bm{a} $ we will use the notation: $a_n = \bm{a} \cdot \bm{\hat{n} } \, ; \,  a_{\perp} = \bm{a} \cdot \bm{\hat{n}_{\perp} }$ and for sake of simplicity we will denote $ V(t) $ simply as $ V $.\\
After one collision at time $ t $ between a particle of position and speed 
$ (\bm{x},\bm{v} )$  and $  C_+(t)  $ the change in momentum along x-axis is (see Appendix)
$ (2m / M ) ( \, v_n -V_n \, )  \bm{\hat{n} }\cdot  \bm{\hat{x} }  $ .

The term $ \Delta V_{coll}(t) $ takes into account all the collisions happening during $ \Delta t $, thus:
\begin{equation}
\label{DP_coll somme in k}
\Delta V_{coll}(t)= \frac{2m}{M}  \sum_{k} ( \, v^k_n -V_n \, )\bm{\hat{n} }\cdot  \bm{\hat{x} }  +\alpha
\end{equation}
where $ k $ labels all particles around the body that are hitting $  C_+(t)  $ within 
$ \Delta t $, and $ \alpha $ denotes terms $ o(\Delta t)  $

Let $ \Delta \bm{x}^i \Delta \bm{v}^j $ be a volume of the phase space of measure
 $ \vert \Delta \bm{x}^i \Delta \bm{v}^j \rvert  = \Delta W $, centered at the point $ (\bm{x}^i,\bm{v}^j)$ and
 $\Delta N (\bm{x}^i,\bm{v}^j,t) $ the number of particles contained in it at time $ t $, so that :
\begin{gather}
\frac{2m}{M} \sum_{k} ( \, v^k_n -V_n \, )\bm{\hat{n} }\cdot  \bm{\hat{x} } =
\frac{2m}{M} \sum_{i j}( \, v^k_n -V_n \, )\bm{\hat{n} }\cdot  \bm{\hat{x} }  \Delta N (\bm{x}^i,\bm{v}^j,t)  \notag \\
= \frac{2}{M} \sum_{i j}( \, v^k_n -V_n \, )\bm{\hat{n} }\cdot  \bm{\hat{x} } \, m \, 
\frac{\Delta N (\bm{x}^i,\bm{v}^j,t) }{\Delta W } \Delta W, \notag
\end{gather}
where $ i, j $  ranges over positions and velocities that will give rise to collision in 
$ (t, t+ \Delta t) $.
At this point letting $ \Delta W \to 0 $ the mean field approximation,whose meaning was mentioned in the introduction, guarantees the  convergence to a finite mass density, i.e :
\begin{equation}
 \lim_{ \Delta W \to 0 } m \, \frac{\Delta N (\bm{x}^i,\bm{v}^j,t) }{\Delta W } = f(\bm x, \bm v,t),
\end{equation}
so that we arrive to:
\begin{equation}
\label{DP coll integrale}
\Delta V_{coll}(t) =\frac{2}{M} \int \limits _{\Omega(\Delta t)}( \, v_n -V_n \, )\bm{\hat{n} }\cdot  \bm{\hat{x} } f(\bm x, \bm v,t) d\bm x d\bm v + \alpha
\end{equation}
where $ \Omega(\Delta t) $ is the $ (\bm x, \bm v)$ region of particles hitting $ C^+(t) $ in $(t, t+ \Delta t)$;\, for further convenience we split this integral into backward  contribution to recoliision 
$  \Omega^L(\Delta t) $ and frontal contribution $ \Omega^R(\Delta t) $:
\begin{equation}
\label{DP coll integrale}
\int \limits _{\Omega^L(\Delta t)}\frac{2}{M}( \, v_n -V_n \, )\bm{\hat{n} }\cdot  \bm{\hat{x} } f(\bm x, \bm v,t) d\bm x d\bm v \,  +
\int \limits _{\Omega^R(\Delta t)}\frac{2}{M}( \, v_n -V_n \, )\bm{\hat{n} }\cdot  \bm{\hat{x} } f(\bm x, \bm v,t) d\bm x d\bm v.
\end{equation}

These region are clearer if written through coordinates moving with the body, namely:
\begin{equation}
\bm{\xi} =\bm{x}- \bm{O}(t)
\end{equation}
through this coordinates we can write  $C^+(t)= \{ (\xi_n, \xi_{\perp}) \in \mathbb{R}^2 : \,  0< \xi_{\perp}< L, \, \xi_n=0 \, \}  $\\
As far as $  \Omega^L(\Delta t) $ is concerned, a collision happens only if $v_n -V_n>0$, besides the particle colliding are those contained in the rectangle: $ -(v_n-V_n) \Delta t<\xi_n < 0$ up to order $ o (\Delta t)$.
In this way we can compute the first integral as:
\begin{gather*}
  \frac{2}{M} \int \limits _{v_n>V_n} d \bm{v }\int \limits _{0}^{L} d\xi_{\perp} 
\int \limits _{-(v_n-V_n) \Delta t}^{0} d\xi_n ( \, v_n -V_n \, )\bm{\hat{n} }\cdot  \bm{\hat{x} } f(\bm  \xi + \bm O(t), \bm v,t)=  \\
=  \frac{2}{M} \Delta t \int \limits _{v_n>V_n} d \bm{v }\int \limits _{0}^{L} d\xi_{\perp} 
( \, v_n -V_n \, )^2 \bm{\hat{n} }\cdot  \bm{\hat{x} } f(\bm  \xi + \bm O(t), \bm v,t) +\alpha
\end{gather*}
In the same way, as far as $  \Omega^R(\Delta t) $ is concerned, a collision happens only if $v_n -V_n<0$, besides the particle colliding are those contained in the rectangle: $ 0<\xi_n < (V_n-v_n) \Delta t$ up to order $ o (\Delta t)$, so that for the second integral it yields:
\begin{gather*}
 \frac{2}{M} \int \limits _{v_n<V_n} d \bm{v }\int \limits _{0}^{L} d\xi_{\perp} 
\int \limits _{0}^{ (V_n-v_n) \Delta t } d\xi_n ( \, v_n -V_n \, )\bm{\hat{n} }\cdot  \bm{\hat{x} } f(\bm  \xi + \bm O(t), \bm v,t)=  \\
= - \frac{2}{M} \Delta t \int \limits _{v_n<V_n} d \bm{v }\int \limits _{0}^{L} d\xi_{\perp} 
( \, v_n -V_n \, )^2 \bm{\hat{n} }\cdot  \bm{\hat{x} } f(\bm  \xi + \bm O(t), \bm v,t) +\alpha
\end{gather*}
In the limit $\Delta t \to 0$ the integrand region becomes $ \{ (\xi_n, \xi_{\perp}): \,  0< \xi_{\perp}< L, \, \xi_n=0 \, \}  $, namely the $C^+ (t) $, which moving back to the previous coordinates reads simply as \footnote{ see \ref{c+-} } $ \{ (x, y) : \, x= X(t) +\eta \cos(\theta) ; \quad y=  \eta \sin(\theta) \, , \, \eta \in [0,L]  \} $, so we finally arrive to:
\begin{gather*}
\begin{split}
\lim_{\Delta t \to 0} \frac{\Delta V_{coll}(t)}{\Delta t} = 
+2 \bm{\hat{n} }\cdot  \bm{\hat{x} } \int \limits _ { 0 }^L  d \eta \int \limits _{ v_n>V_n } d \bm{v}
( \, v_n -V_n \, )^2  f_-( X(t) +\eta \cos(\theta),\eta \sin(\theta) ,\bm{v},t)  +\\
- 2 \bm{\hat{n} }\cdot  \bm{\hat{x} } \int \limits _ { 0 }^L d \eta \int \limits _{ v_n<V_n } d \bm{v}
( \, v_n -V_n \, )^2  f_-(X(t) +\eta \cos(\theta),\eta \sin(\theta) ,\bm{v},t) 
\end{split}
\end{gather*}
as we stated earlier the contribution to $F(t)$ from the whole body is twice that on $C^+(t)$, besides noticing that $\bm{\hat{n} }\cdot  \bm{\hat{x} }= \sin (\theta)$ and $v_n -V_n = \bm v ' \cdot \bm{\hat{n} }$ concludes the derivation of eq.(\ref{d/dt V = E- F(t)}) and (\ref{F}).\\

Now, thanks to boundary conditions we can compute the friction in terms of the 
initial mass density. 
Indeed let $ \bm{x}(s,t,\bm{x},\bm{v}) $, $ \bm{v}(s,t,\bm{x},\bm{v}) $ be the position and velocity of a particle at time $ s \leq t $, that at time  $ t $ occupies position  $ \bm{x} $ and velocity
$\bm{v} $ ; 
conservation of mass imply that the P.d.f. stays constant along particles trajectories and in particular 
\begin{equation}
f(\bm{x}, \bm{v},t)= f_0( \bm{x}(0,t,\bm{x},\bm{v}) , \bm{v}(0,t,\bm{x},\bm{v}))
\end{equation}
so that the problem of finding the gas distribution reduces to that of tracking the particles trajectories.

Given the evolution of the body $X(t) $, $ V(t) $, there is a unique backward time evolution leading to the initial position and velocity, (see Figure 1). 
Such backward evolution is free motion up to possible collision-times in which the particle hits the body. On these times we keep track of  the particle displacement through condition (\ref{boundary conditions}). We proceed in this way until we reach the desired
 $\bm{x}(0,t,\bm{x},\bm{v}) $ ,  $ \bm{v}(0,t,\bm{x},\bm{v})$.
At the end using the initial state of the gas distribution, eq.(\ref{distribuzione di Maxwell}), we obtain

\begin{equation}
\label{F(t)-Angle}
\frac{F(t)}{ 4k \sin(\theta)}=  \int \limits _ { 0 }^L  d \eta \int \limits _{ \bm{v}' \cdot \bm{\hat{n}}< 0 } d \bm{v}
( \bm{v}' \cdot \bm{\hat{n}}  )^2  e^{-\beta \bm{v}_0^2}\\
-  \int \limits _ { 0 }^L d \eta \int \limits _{ \bm{v}' \cdot \bm{\hat{n}} > 0 } d \bm{v}
( \bm{v}' \cdot \bm{\hat{n}}  )^2  e^{-\beta \bm{v}_0^2}
\end{equation}
where $ \bm{v}_0 =\bm{v}(0, t, X(t) +\eta \cos(\theta), \eta \sin(\theta), \bm{v}) $ and $ k= \rho ( \frac{\beta}{\pi} ) $. \\
Note that in order to compute $ F(t) $ we need to evaluate $ \bm{v}_{\zeroped{0}} $ and hence to know all the previous history  $\{ X(s), V(s) , s<t \}$, on the other hand, if the light particle goes back without undergoing any collision, then $ \bm{v}_0=\bm{v}$ and the friction term is easily computed\\
Note also how the limit case '' $ \theta \rightarrow \frac{\pi}{2} $ '', which gives back the normal stick in two dimensions, brings back the usual friction term, it suffices to note that under this limit
$  \bm{ v }' \cdot \bm{\hat{n} } = v_x -V(t)  $.\\

\section{ Estimate of $ F^V(t) $ }

In this section we will study the main problem, namely the friction exerted on a body fixed at (any) constant velocity.

We start computing $  F^V(t)  $, the friction term for the body $ C(t) $ moving at a constant velocity, $ V(t) \equiv V $,
in this case  there can't be recollisions on the back of the body, though there can be on the front of it \footnote{This is indeed the main difference with all previous cases for constant velocities}, therefore eq. (\ref{F(t)-Angle}), in the case of a constant velocity, gets
\begin{equation}
\frac{F^V(t)}{ 4k \sin(\theta)}=  \int \limits _ { 0 }^L  d \eta \int \limits _{ \bm{v}' \cdot \bm{\hat{n}}< 0 } d \bm{v}
( \bm{v}' \cdot \bm{n}  )^2  e^{-\beta \bm{v}_0^2}\\
-  \int \limits _ { 0 }^L d \eta \int \limits _{ \bm{v}' \cdot \bm{\hat{n}} > 0 } d \bm{v}
( \bm{v}' \cdot \bm{\hat{n}}  )^2  e^{-\beta \bm{v}^2}
\end{equation}
with $ \bm{v}'  = \bm{v} -V\bm{\hat{x}}  $, for further convenience we will write 
$ \bm{ V} =V\bm{\hat{x}}   $.

In the first integral there is a subset of the region $ \bm{v}' \cdot \bm{\hat{n}}< 0  $ 
 such that $ \bm{v}_0= \bm{v} $, that is particle not coming from a previous recollision, and a subset such that $ \bm{v}_0 \neq \bm{v} $, that is particles coming from a previous recollision, for these ones we now compute  $ \bm{v}_0  $.

As $ \bm{\hat{n}}   $ is the normal to the right side of $ C_+(t) $,
we write $ \bm{\hat{p}}  $ as the normal to the right side of $ C_-(t) $ ,  $ \bm{\hat{p} }  =( \sin(\theta), \cos(\theta ) ) $ and
$ \bm{\hat{p}_{\perp}} = ( -\cos(\theta), \sin(\theta ) ) $ as  its upward normal; besides  for any vector $ \bm{w} $ we call $ w_p = \bm{w} \cdot \bm{\hat{p}}  $ and 
 $ \bm{w}_{p_{\perp}}= \bm{w} \cdot \bm{\hat{p}_{\perp}} $.\\
Having reduced to the  $ C_+(t) $ in the integral, particles coming from a previous collision can only come from  $ C_-(t) $; besides the condition $ \theta \in [ \frac{\pi}{4}, \frac{\pi}{2}) $ ensure at most one recollision for constant velocities of the body.
In the reference system of the body the computation  reduces to an elastic collision between a fixed tilted wall and a particle with velocity  $ \bm{v}_0' =  \bm{v}_0-  \bm{V} $ before and 
$ \bm{v}' = \bm{v} -\bm{ V}$ after the collision.
We thus arrive to
\begin{subequations}
\label{urti}
\begin{gather}
v_{0p} = 2 V_p -v_p \\
v_{0p_{\perp}} =v_{p_{\perp}}
\end{gather}
\end{subequations}
This allow us to compute 
\begin{equation}
\label{v0quadro}
\bm{v}_0^2= \bm{v}^2 -4V_p (\bm{v} -\bm{V})\cdot \bm{\hat{p}}
\end{equation}
we stress again that this  is the initial velocity of particles that underwent previous collision with $ C_-(t) $, we now compute the exact region in phase space that leads to recollisions.

Let $ \bm{R}(\eta)= ( X(t) +\eta \cos(\theta) , \, \eta \sin(\theta) ) $ be a point of $ C_+(t) $ and $ \bm{Q}= ( X(t) +L \cos(\theta) , \, -L sin(\theta) ) $ be the extreme lower point of 
$ C_-(t) $, then 
\begin{equation}
 \hat{ \bm{\psi}  }(\eta) =N ( (\eta +L)\sin(\theta), (L- \eta)\cos(\theta) ) 
\end{equation}
is the vector such that 
\begin{subequations}
\begin{gather}
\hat{ \bm{\psi}  }(\eta) \cdot \hat{ \bm{RQ}  } (\eta)=0 \\
\hat{ \bm{\psi}  }(\eta) \cdot \hat{ \bm{n}  } >0
\end{gather}
\end{subequations}
and it is clear that the subset leading to recollision is within the region\footnote{ From now on we will simply write $ \hat{ \bm{\psi}  }  $} 
 $ \{   \bm{v} \in \mathbb{R}^2 : \bm{v}'  \cdot \hat{ \bm{n}  } <0  $ , $  \bm{v}'  \cdot \hat{ \bm{\psi}  } > 0  \}$,
 where $ \bm{v}' = \bm{v} -\bm{V} $.\\
Now, to define the exact recollision region we follow backwards in time a particle hitting  
$ \bm{R}(\eta) $ with velocity $ \bm{v} $ at time $ t $.
This particle will hit $C_-(t)  $ if it covers in the $ \bm{\hat{p}} $ direction, a lenght greater than $ \bm{OR} \cdot \bm{\hat{p}} = \eta \sin(\alpha) $ where $ \alpha =\pi -2\theta $, 
 which is to say 
$   (\bm{v}'  \cdot \bm{\hat{p}} ) t \geq \eta \sin(2\theta) $.
\\
Finally the exact recollision region reads
\begin{equation}
\mathcal{R }_{\theta}(\eta,t)= \{  \bm{v} \in \mathbb{R}^2 : \bm{v}'  \cdot \hat{ \bm{n}  } <0   , \ \bm{v}'  \cdot \hat{ \bm{\psi}  } > 0 , \  \bm{v}'  \cdot \bm{\hat{p}}  \geq \frac{\eta}{t}  \sin(2\theta)  \}
\end{equation}
Before proceeding we note that we can rewrite $ F^V(t) $ in a convenient way: 
 \begin{equation}
 F^V(t) = F_{0}(V)+g(V,t)
 \end{equation}
with
\begin{gather}
\frac{F_0(V)}{ 4k \sin(\theta)}=  \int \limits _ { 0 }^L  d \eta \int \limits _{ \bm{v}' \cdot \bm{\hat{n}}< 0 } d \bm{v}
( \bm{v}' \cdot \bm{n}  )^2  e^{-\beta \bm{v}^2}
-  \int \limits _ { 0 }^L d \eta \int \limits _{ \bm{v}' \cdot \bm{\hat{n}} > 0 } d \bm{v}
( \bm{v}' \cdot \bm{\hat{n}}  )^2  e^{-\beta \bm{v}^2}\\
\frac{g(V,t)}{4k \sin(\theta)} =  \int \limits _ { 0 }^L  d \eta \int \limits _{ \bm{v}' \cdot \bm{\hat{n}}< 0 } d \bm{v}
( \bm{v}' \cdot \bm{n}  )^2  \huge ( e^{-\beta \bm{v}_0^2}- e^{-\beta \bm{v}^2} \huge ).
\end{gather}

An important and new feature of this concave case is that even with constant velocity there is a time dependent term in the friction .
Notice in fact that $ \bm{v}_0 (\bm{x},\bm{v},t )= \bm{v} $ for $ \bm{v} \notin  \mathcal{R }_{\theta}(\eta,t) $ hence leading to a null contribution in $ g(V,t) $,
 therefore we arrive to

\begin{equation}
g(V,t)= 4k \sin(\theta) \int \limits _ { 0 } ^L  d \eta \int \limits _{ \mathcal{R }_{\theta}(\eta,t)} d \bm{v}'
( \bm{v}' \cdot \bm{n}  )^2   e^{-\beta (\bm{v}'+ \bm{V} )^2} [ e^{4\beta V_p (\bm{v}' \cdot \bm{\hat{p}} ) }  -1  ].
\end{equation}

Where we used eq.(\ref{v0quadro}) because  the integral is over the exact recollision region. \footnote{ Moreover since $  \bm{v}' =\bm{v} -\bm{V} $ \, it yields $  d \bm{v}' = d \bm{v} $. } 
In this region particles had a previous collision thus $ v_{0p}<V_p $ and through eq. (\ref{urti}) it follows that $ \bm{v}'  \cdot \bm{\hat{p}}  \geq 0 $ .\\
Moreover it is clear that $ \mathcal{R }_{\theta}(\eta,t) \subseteq \mathcal{R }^{\infty}_{\theta}(\eta)  $ where 
\begin{equation}
\mathcal{R }^{\infty}_{\theta}(\eta)= \{  \bm{v} \in \mathbb{R}^2 : \bm{v}'  \cdot \hat{ \bm{n}  } <0   , \ \bm{v}'  \cdot \hat{ \bm{\psi}  } > 0  \},
\end{equation}
this can be looked at as the recollision region for $ t \rightarrow \infty $; these relations imply 
\begin{gather}
0 \leq g(V,t) \leq g^{\infty}(V)\\
g^\infty(V)= 4k \sin(\theta)\int \limits _ { 0 } ^L  d \eta \int \limits _{ \mathcal{R }_{\theta}^{\infty}(\eta)} d \bm{v}' (\bm{v}' \cdot \bm{n}  )^2   e^{-\beta (\bm{v}'+ \bm{V} )^2} [ e^{4\beta V_p (\bm{v}' \cdot \bm{\hat{p}} ) }  -1  ]
\end{gather}
this means that the concavity add a positive contribution $ g(V,t) $ to the friction and it is limited in time.

In order to study the time dependence we write  $ g(V,t)= g^{\infty}(V) -\Delta g(V,t) $ with :
\begin{gather}
\label{DG}
\Delta g(V,t) =  g^\infty(V)- g(V,t) = 4k \sin(\theta)\int \limits _ { 0 } ^L  d \eta \int \limits _{ \Delta\mathcal{ R }(t)} d \bm{v}'
( \bm{v}' \cdot \hat{ \bm{n}  }  )^2   e^{-\beta (\bm{v}'+ \bm{V} )^2} [ e^{4\beta V_p (\bm{v}' \cdot \bm{\hat{p}} ) }  -1  ]\\
\Delta\mathcal{ R }(t) =\mathcal{R }_{\theta}^{\infty}(\eta) \diagdown \mathcal{R }_{\theta}(\eta,t) =  \{  \bm{v} \in \mathbb{R}^2 : \bm{v}'  \cdot \hat{ \bm{n}  } <0   , \ \bm{v}'  \cdot \hat{ \bm{\psi}  } > 0 , \ 0 \leq \bm{v}'  \cdot \bm{\hat{p}}  \leq \frac{\eta}{t}  \sin(2\theta)  \}
\end{gather}
We now perform the following rotation in the velocity space in order to normalize the integral region:
\begin{gather}
\bm{v}' = \hat{R}(\phi) \bm{w} \\
\hat{R}(\phi) =
\begin{pmatrix}
 \cos(\phi) & -\sin(\phi) \\
 \sin(\phi) &  \cos(\phi) 
\end{pmatrix}
\end{gather}
where $ \phi = \frac{\pi}{2}- \theta $.
We note that scalar products keep the same structure under $\hat{R}$, that is for every vector
$ \bm{q} $ it yields 
\begin{equation}
\bm{v}' \cdot \bm{q} = \hat{R}(\phi) \bm{w} \cdot \hat{R}(\phi) \bm{q}_w = \bm{w} \cdot  \bm{q}_w ,
\end{equation}
where $ \bm{q}_w = \hat{R}(-\phi) \bm{q}  $; 
note also  that $ \bm{\hat{p}} = (\cos(\phi), \sin(\phi)) $.
Through this change of variables, for which $ d \bm{w}= d \bm{v}' $,  we arrive to:
 \begin{gather}
\Delta g(V,T)=  4k \sin(\theta) e^{-\beta \bm{V}^2}\int \limits _ { 0 } ^L  d \eta \int \limits _{ \Delta\mathcal{ R }_w(T)} d \bm{w}
( \bm{w} \cdot \bm{n}_w  )^2   e^{-\beta (\bm{w}^2+ 2\bm{w} \cdot \bm{V}_w )} [ e^{ 4\beta V_p (w_1) }  -1  ]\\
\Delta\mathcal{ R }_w(T) = \{  \bm{w} \in \mathbb{R}^2 : \bm{w} \cdot \hat{ \bm{n} }_w <0   , \ \bm{w}  \cdot \hat{ \bm{\psi} }_w > 0 , \ 0 \leq  w_1  \leq \ T\eta  \}
\end{gather}
where $  \hat{ \bm{\psi}  }_w(\eta) =N_w ( L- \eta \cos(2\theta),  -\eta \sin(2\theta) ) $, 
$ \hat{ \bm{n}  }_w =(- \cos(2\theta), -\sin(2\theta) ) $ and
$ T= \frac{sin(2\theta)}{t} $,
besides $ \bm{w} \cdot \hat{ \bm{n} }_w < 0  $ can be rewritten as $ w_1 n_{w_{1}} +w_2 n_{w_{2}} < 0  $ and, keeping in mind that 
$ n_{w_{2}} < 0 $ this leads to $ |\frac{n_{w_{1}} }{n_{w_{2}} } | w_1 < w_2 $,  similarly for $ \bm{w} \cdot \hat{ \bm{\psi} }_w > 0  $, leading us to :
 \begin{gather}
 \label{Deata grande}
\Delta g(V,T)  =4k \sin(\theta) e^{-\beta \bm{V}^2} \int \limits _ { 0 } ^L  d \eta \int \limits _{0}^{T \eta} d w_1  \int \limits _{a w_1}^{b w_1} d w_2 f(w_1,w_{2})\\
f(w_1,w_{2})= ( \bm{w} \cdot \bm{n}_w  )^2   e^{-\beta (\bm{w}^2+ 2\bm{w} \cdot \bm{V}_w )} [ e^{ 4\beta V_p (w_1) }  -1  ]
 \end{gather}
 where 
\begin{gather}
 a= |\frac{n_{w_{1}} }{n_{w_{2}} } | = \tan(2\theta -\frac{\pi}{2}) \\
\label{b}
 b= |\frac{\psi_{w_{1}} }{\psi_{\psi_{2}} } | = \frac{L- \eta \cos(2\theta)}{\eta \sin(2\theta)}
\end{gather} 
we incidentally observe that $ b \geq  \tan(\theta) $, which we will be of use later.\\

We now give an exact estimate in time of $ \Delta g(V,T) $.
In the sequel  $ C $ and $ C_n $ will denote positive constants possibly depending on 
$  \beta $, $ \theta  $, $ V $ and  $ L $. \\
From eq. (\ref{DG}), using that $ e^{-\beta (\bm{v}'+ \bm{V} )^2} \leq 1 $ , 
$ (\bm{v}'  \cdot \hat{ \bm{n}  })^2 \leq |\bm{v}' |^2$ and passing to the  variables 
''$ \bm{w} $'' we arrive to
\begin{equation}
\Delta g(V,t) \leq C \int \limits _ { 0 } ^L  d \eta \int \limits _{ \Delta\mathcal{ R }_w} d \bm{w} \ \bm{w} ^2 [ e^{C_1 w_1 } -1],
\end{equation}
clearly $ w_1 \leq T\eta $ and $ \bm{w} ^2  \leq |(T \eta, b T \eta)|^2 \leq C T^2 $ since, from (\ref{b}), $ b\eta $ is bounded, moreover  $ |\Delta\mathcal{ R }_w| \leq C  T^2 $, this leads us to
\begin{equation}
\Delta g(V,T) \leq C T^4 (e^{C_1 T} -1)
\end{equation} 
and then 
\begin{equation}
 \lim_{ T \to 0 } \frac{\Delta g(V,T)}{T^5} \leq C.
\end{equation}
For the lower bound we start again from eq. (\ref{DG}) and use that $ |\bm{v}' +\bm{V} |^2 \leq (|\bm{v}'|+|\bm{V} | )^2 $, leading us to
\begin{equation}
\Delta g(V,t) \geq C \int \limits _ { L/2 } ^L  d \eta \int \limits _{ \Delta\mathcal{ R }(t)} d \bm{v}'
( \bm{v}' \cdot \hat{ \bm{n}  }  )^2   e^{-\beta (|\bm{v}'|^2+2|\bm{v}'| |\bm{V}| )} [ e^{4\beta V_p (\bm{v}' \cdot \bm{\hat{p}} ) }  -1  ]
\end{equation}\\
Now passing to the variables ''$ \bm{w} $'' and using again that $ \bm{w} ^2  \leq C T^2 $, \\ we have that $ \bm{w}^2 +2|\bm{w}| V \leq C T^2 + C_1T  \leq C_2 T $, the last inequality holding for $ T $ small enough;  we thus arrive to
\begin{gather}
\Delta g(V,t) \geq C \int \limits _ { L/2 } ^L  d \eta \int \limits _{0}^{T \eta} d w_1  \int \limits _{a \ w_1}^{b \ w_1} d w_2 
( \bm{w} \cdot \hat{ \bm{n}  }_w  )^2   e^{-C_1 T} [ e^{ C_2 w_1 }  -1  ]  \geq\\
\label{last}
 \geq C \int \limits _ { L/2 } ^L  d \eta \int \limits _{T \eta /2}^{T \eta} d w_1  \int \limits _{a_{\epsilon} \  w_1}^{b \ w_1} d w_2 
\ T^2   e^{-C_1 T} [ e^{ C_2 T }  -1  ]  \geq \\
 \geq C T^4  e^{-C_1 T} [ e^{ C_2 T }  -1  ] ,
\end{gather}
in the last inequalities we shrunk the integral region in: $ a_{\epsilon} w_1 \leq w_2 \leq b w_1 $, where \\ $ a_{\epsilon}= \tan(2\theta -\pi/2 +\epsilon)  $ can be chosen satisfying: $ \tan(2\theta -\pi/2)=a<a_{\epsilon} < \tan(\theta) <b  $, with $ 0 <\epsilon < \pi/2- \theta  $; in this region it holds:
\begin{equation}
( \bm{w} \cdot \hat{ \bm{n}  }_w  ) = |n_{w_2}|(a w_1 -w_2) \leq  -|n_{w_2}|(a_{\epsilon}- a)w_1 < -C T 
\end{equation}
leading to  $ ( \bm{w} \cdot \hat{ \bm{n}  }_w  )^2 \geq C T^2 $ which was used in (\ref{last}). 
Therefore we arrived to 
\begin{equation}
 \lim_{ T \to 0 } \frac{\Delta g(V,T)}{T^5} \geq C .
\end{equation}
Now, changing  back to $ t= \frac{\sin(2\theta)}{T} $, together with the fact that $ \Delta g(V,t)  $ is bounded, leads us to the desired estimate for $ \Delta g(V,t) $, namely
\begin{equation}
\frac{C_l}{(1+t)^5} \leq \Delta g(V,t) \leq \frac{C_u}{(1+t)^5} \quad \forall t \geq 0 
\end{equation}
and finally 
\begin{equation}
g^{\infty}(V) - \frac{C_u}{(1+t)^5} \leq  F^V(t) -F_{0}(V) \leq g^{\infty}(V) - \frac{C_l}{(1+t)^5} \quad \forall t \geq 0.
\end{equation} 

\section{Absence of a stationary solution}

We now show how having computed the fricion term in the previous problem also allow us to prove that the actual dynamic (\ref{d/dt V = E- F(t)}) of such a concave body  (no more fixed at any constant velocity) has no stationary solution.
For the previous cases of the disk, the convex body and the cylinder there is a stationary solution $ V(t) \equiv \bar{V}   $ that can be found directly from eq. (\ref{d/dt V = E- F(t)}) namely :
\begin{gather}
0= E - F^{ \bar{V} }(t), \quad  \forall t \geq 0 \\
V(0) = \bar{V}
\end{gather}
where $F^{ \bar{V} }(t) $ is the friction term computed for the constant velocity $\bar{V}  $;
in those cases (see \cite{nostro}, \cite{E=0} and \cite{CILINDRO}) it yields  $F^{ \bar{V} }(t) =F_0(\bar{V}) \ \forall t\geq 0 $ because recolision terms are absent for constant velocities, the stationary solution is therefore  the one such that $ E - F_0 (\bar{V}) =0  $
which was called $ V_{\infty} $ and was also the limiting velocity of the body.\\
In the present case of concavity  the possible stationary velocity has to be such that
\begin{gather}
\label{stationary-angle}
0= E - F^{ \bar{V} }(t)= E-F_{0}(\bar{V})- g(\bar{V},t), \quad  \forall t \geq 0 \\
V(0) = \bar{V}.
\end{gather}
This implies the following necessary condition on the stationary velocity: 
\begin{equation}
 \frac{d g(\bar{V},t)  }{d t} = 0 \quad \forall t \geq 0 
\end{equation}
and through the previous analysis we can compute this derivative from (\ref{Deata grande}) in the following way (we use for our convenience the variable employed earlier $ T= \frac{sin(2\theta)}{t} $ ):
\begin{equation}
- \frac{d g(V,T)}{dT}=  \frac{d \Delta g(V,T)}{dT}= 4k \sin(\theta) e^{-\beta \bm{V}^2} \int \limits _ { 0 } ^L  d \eta \ \eta \int \limits _{a \eta T}^{b \eta T} d w_2 f(\eta T, w_2)
\end{equation}
We immediately observe that (\ref{stationary-angle}) is not satisfied by $ \bar{V} =0 $
since $ F_{0}(0)= g(0,t)=0  $, while for  $ \bar{V} >0 $ we observe by immediate inspection  that $ f(\eta T, w_2) >0 \ \forall \bar{{V}} >0 $, thus 
\begin{equation}
 \frac{d g(\bar{V},T)}{dT} <0 \quad \forall \bar{{V}} >0 ,T >0 
\end{equation}
which exclude the possibility of a stationary solution. 

\section{Comments}

In this work we studied what we believe to be the main feature of general kinds of concavities in their interaction with a fluid, namely tilted walls that correlate the body with itself via the bouncing of gas particles, we also showed how this affects even the simple friction for constant velocities, in particular the presence of a time dependence in it.   
As it was reasonable to expect, not only this contribution is bounded but also vanishing in time trough the estimate we computed.\\
We studied the two dimensional case in order to focus on the main feature of the problem,
avoiding long technical details; following the reasoning of the present work it is evident that the only tweak between two and three dimensions is the exact power at which the time dependence vanishes, the rest being the same. 

We also stress that these results, as those of the main articles to which we refer,  strongly
depend on the choice of a perfect gas as the mean exerting friction, recollisions in fact come essentially from the long memory effect of gas particles in the setting of a Vlasov gas.\\
We end this section observing  that the absence of a stationary solution doesn't exclude a limiting velocity in the dynamic, which is instead expected, in fact we showed that the time dependence of this new friction term is vanishing reasonably fast in time and through heuristic reasoning we expect  the limit velocity to be a $ \bar{V}_{\infty}  $ satisfying 
\begin{equation}
E-F_{0}(\bar{V}_{\infty}) -g^{\infty}(\bar{V}_{\infty}) =0
\end{equation}
which is  a correction of the previous $ V_{\infty} $ ( see \cite{E=0}, \cite{CONVEX} and \cite{CILINDRO} ) due to the non vanishing component of the concavity friction, thus depending also on the tilting parameter 
$ \theta $, in particular, as it is clear from the foregoing analysis, $ \bar{V}_{\infty} \to V_{\infty} $ for $ \theta \to \pi / 2 $.\\
The actual dynamic for this kind of general concavity is still to be studied, but this work together with \cite{CILINDRO} show how the shape of the body, in particular a meaningful concavity, is essential in the interaction with the medium it is moving in.
\\

\appendix
\section{Appendix}

We derive here collision conditions (2.3) and (2.4). In what follows we denote by $M$ and $\bm V$ the mass and velocity of the body and by $m$ and $v$ mass and velocity of a particle which will be assumed to collide elastically with the body.\\
We consider that the particle hits the body at $ \bm x \in  C(t)  $ at time $ t $ and define its pre(post)collisional velocity as $ \bm{v} $ $  ( \bm{v}' )$, projecting conservation of momentum equation on $ \bm{\hat N} $ and $\bm{\hat N}_{\perp} $ direction, together with the conservation of  kinetic energy imply for the body:
\begin{gather}
V'_N=V_N+\frac{2m}{M+m}(v_N-V_N)   \\ 
V_{N_{\perp}}'= V_{N_{\perp}}
\end{gather} 
where $V'_N$ and $v'_N$ are post-collisional velocities along the $ \bm{\hat N} $ direction.\\
While for the particle:
\begin{gather}
v_N'=V_N-\frac{M-m}{M+m}(v_N-V_N) \\
v_{N_{\perp}}'= v_{N_{\perp}}
\end{gather} 

Being $M>>m$, we have

\begin{subequations}
\begin{gather}
v_N'\simeq 2V_N-v_N \\
v_{N_{\perp}}'= v_{N_{\perp}}.
\end{gather} 
\end{subequations}
and for the body:
\begin{gather}
V_N' \simeq V_N+\frac{2m}{M}(v_N-V_N)  
\end{gather}
Therefore 
\begin{equation}
\bm V' -\bm V = (V_N'- V_N) \bm{\hat{N}} = \frac{2m}{M}(v_N-V_N) \bm{\hat{N}}
\end{equation}
 
hence the momentum change along the $\bm{\hat{x}} $ direction is
\begin{equation}
\frac{2m}{M}(v_N-V_N) \bm{\hat{N}} \cdot \bm{\hat{x}}.
\end{equation}


\begin{thebibliography}{99}


\bibitem{diffusivo}
K. Aoki, G. Cavallaro, C. Marchioro, and M. Pulvirenti, 
\emph{ On the motion of a body in thermal equilibrium immersed in a perfect gas},
Math. Model. Numer. Anal. $ \bm{42}  $, 263-275 (2008).

\bibitem{NUMERICO 2}
K. Aoki, T. Tsuji, and G. Cavallaro, \emph{Approach to steady motion of a
plate moving in a free-molecular gas under a constant external force },
Phys. Rev. E $ \bm{80}  $, 016309, 1-13 (2009).

\bibitem{nostro}
S. Caprino, C. Marchioro, and M. Pulvirenti, \emph{Approach to equilibrium 
in a microscopic model of friction},
Comm. Math. Phys. $ \bm{264}  $ , 167-189 (2006).

\bibitem{E=0}
S. Caprino, G. Cavallaro, and C. Marchioro, \emph{On a microscopic
model of viscous friction}, Math. Models Methods Appl. Sci. $ \bm{17} $, 1369-1403 (2007).

\bibitem{CONVEX}
G. Cavallaro,	\emph{On the motion of a convex body interacting with a perfect
gas in the mean-field approximation}, Rend. Mat. Appl. $ \bm{27 } $, 123-145 (2007).

\bibitem{strauss}
X. Chen, W. Strauss, 
\emph{ Approach to Equilibrium of a Body Colliding Specularly and Diffusely with a Sea of Particles},
Arch. Rational Mech. Anal.  (2013), Anal. $ \bm{211 } $, 879-910  (2014). 

\bibitem{CERCIGNANI}
C. Cercignani,	\emph{The Boltzmann Equation and Its Applications}, 
Applied Mathematical Science $ \bm{67} $, New York: Springer Verlag, (1988).

\bibitem{BLA}
R.L. Dobrushin, \emph{Vlasov equations}, Sov. J. Funct. Anal. $ \bm{13} $, 115-123
(1979).




\bibitem{altro}
G. Cavallaro,	 C. Marchioro, \emph{On the motion of an elastic body in a free gas}, 
Reports in Mathematical Physics $ \bm{69 } $,  251-264 (2012).

\bibitem{CILINDRO}
F. Sisti,	 C. Ricciuti, \emph{Effects of concavity on the motion of a body immersed in a Vlasov gas}, 
SIAM Journal on Mathematical Analysis, (2014),  In Press.

\bibitem{NUMERICO 1}
T. Tsuji and K. Aoki, \emph{Decay of a linear pendulum in a free-molecular
gas and a special Lorentz gas, }  J. Stat. Phys. $ \bm{146(3)}$  , 620-645 (2012).






\end{thebibliography}
\end{document}